\documentclass[twocolumn,aps,superscriptaddress]{revtex4-1}
\usepackage{amsfonts, relsize, color}
\usepackage{graphics}
\usepackage{graphicx}
\usepackage{subfig}
\usepackage{hyperref}
\usepackage{amssymb}
\usepackage{mathrsfs}
\usepackage{siunitx}
\usepackage{enumitem}
\usepackage{braket}
\usepackage{mathtools}
\usepackage{pgfplots}

\usepackage{diagbox}
\usepackage{bm}

\newcommand{\be}{\begin{equation}}
\newcommand{\ee}{\end{equation}}

\DeclareMathOperator\erf{erf}
\begin{document}
\title{Mass fluctuations and absorption rates in Dirac materials sensors}
\renewcommand{\andname}{\ignorespaces}
\author{Bart Olsthoorn}
\email{bartol@kth.se}
\affiliation{NORDITA, KTH Royal Institute of Technology and Stockholm University, Roslagstullsbacken 23, SE-106 91 Stockholm, Sweden}
\author{Alexander V. Balatsky}
\email{avb@nordita.org}
\affiliation{NORDITA, KTH Royal Institute of Technology and Stockholm University, Roslagstullsbacken 23, SE-106 91 Stockholm, Sweden}
\affiliation{Department of Physics, University of Connecticut, Storrs, CT 06269, USA}

\date{\today}

\begin{abstract}
We study the mass fluctuations in gapped Dirac materials by treating the mass-term as both a continuous and discrete random variable. Gapped Dirac materials were proposed to be used as materials for Dark matter sensors. One thus would need to estimate the role of disorder and fluctuations on the interband absorption of dark matter. We find that both continuous and discrete fluctuations across the sample introduce tails (e.g. Lifshitz tails) in the density of states and the interband absorption rate. We estimate the strength of the gap filling and discuss implications of these fluctuations on the performance as sensors for Dark matter detection. The approach used in this work provides a basic framework to model the disorder by any arbitrary mechanism on the interband absorption of Dirac material sensors.
\end{abstract}

\maketitle
\section{Introduction}
A large number of (mostly inorganic) three-dimensional Dirac and Weyl semimetals have been discovered recently~\cite{doi:10.1080/00018732.2014.927109,RevModPhys.90.015001,Xia2009,Tanaka2012,PhysRevX.5.031013}. Prominent examples of materials with a linear crossing of energy bands are graphene and topological insulators. The density of states of a $d$-dimensional Dirac material (DM) scales as $\nu(E)\sim|E|^{d-1}$ and vanishes at the Dirac point. 

These Dirac or nodal points are present due to symmetry of the system: e.g. Dirac nodes are either present at high symmetry points due to crystalline symmetry or can emerge as accidental crossings at arbitrary points in the Brillouin zone being protected by the topology of the band structure~\cite{PhysRevB.95.041103,doi:10.1080/00018732.2014.927109,PhysRevB.94.155108}.

\begin{figure}[b]
    \centering
    \includegraphics{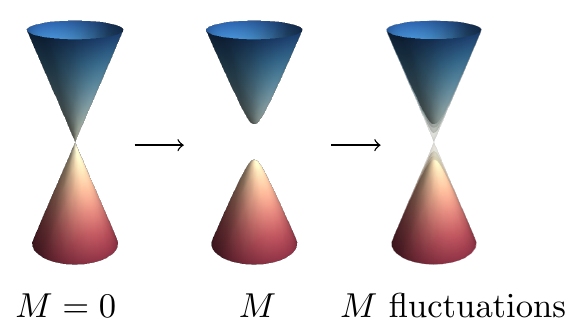}
    \caption{Three stages to arrive at mass fluctuations in a Dirac cone. The gap is induced by the $\sigma_z$-term and different realizations of $M$ are averaged over.}
    \label{fig:cones}
\end{figure}

Electronic states at the Dirac point are thus highly tunable in the presence of the symmetry breaking fields that control the topology of nodal states and open up the gap in the nodal point, see Fig~\ref{fig:cones}. Small symmetry breaking terms and small magnetic fields will control  the  gap energy scale and can be made orders of magnitude smaller than typical semiconducting energy gaps.  Gapped Dirac materials with small gaps (gaps on the order of 10-100 meV)  will have a range of applications.  In the case of graphene, an energy gap can be introduced by adding a mass term that breaks the sublattice symmetry  \cite{doi:10.1080/00018732.2014.927109} and gapped graphene will have applications in optics and electronics. Similarly, a gapped Quantum anomalous Hall  state of massed topological insulator surface states with persistent edge currents forms in the presence of the polarized spins  \cite{doi:10.1146/annurev-conmatphys-031115-011417}. 
 Another interesting example is the recent proposal to use Dirac materials for Dark matter  detection~\cite{PhysRevD.97.015004, doi:10.1002/pssr.201800293}, as the material for Dark matter sensors, see Section~\ref{sec:gapped_dm_for_dm}. 

In the typical case of doped semiconductors where the gap is controlled by dopants, impurities will have detrimental effects due to disorder and fluctuations of impurity distribution aside from primary role in gap modifications. The role of impurity fluctuations on the electronic structure was studied using effective models going back to earlier literature ~\cite{lifshitz1988introduction,Bassani_1974,RevModPhys.64.755}.

Alternatively, one can use \textit{ab initio} methods like density functional theory (DFT) to estimate changes in the electronic structure~\cite{RevModPhys.86.253, PhysRevB.89.075103}. More recently, new methods using high-throughput DFT studies with machine learning were also  used to predict the impurity properties of materials~\cite{2019arXiv190602244M}.

The majority of the studies to date have been focused on conventional materials with parabolic dispersions. Investigations of the role of disorder in Dirac and Weyl materials is more recent and is still ongoing \cite{PhysRevX.6.021042,PhysRevB.89.245110,doi:10.1146/annurev-conmatphys-033117-054037,PhysRevB.96.201401}.  Most of the discussion to date was centered on the role of disorder on gapless Dirac states. It is equally important to elaborate on the effects of disorder and fluctuations on the gapped Dirac material. In this paper we discuss the role of impurity fluctuations on the gapped Dirac materials. We find that the effects of disorder and fluctuations will be pronounced in case of small gaps in Dirac materials and specifically we find:

1) To estimate the effects of disorder we  consider the role of a random mass term on the gapped Dirac materials. There are many physical mechanisms for opening the gap, such as, strain, dopants and the induced magnetic field due to spins. Any of the mechanisms mentioned are subject to disorder and fluctuations. For example, strain cannot be constant throughout the sample. Dopants cannot be placed to such a degree that the resulting structure is homogeneous. Any spin texture will produce an inhomogeneous $m$-term in the Dirac equation. To capture any of these mechanisms in our model we consider a randomly distributed $m$-term.

2) We find that both the discrete and continuous random mass-term induce a soft gap in the spectrum and lead to a much smaller effective gap. The density of states of our model with a normally distributed $m$-term vanishes quadratically at the Dirac point but show tails deep in the gap. The presence of rare region  (i.e. Lifshitz tails~\cite{0038-5670-7-4-R03}) causes the density of states to remain finite close to Dirac point, i.e. in the regions with large dopant concentration, we find impurity band tails extended all the way to zero energy. 

3) The presence of gap filling fluctuations in sensor materials leads to stringent constraints on how small the gap should be and what level of gap inducing disorder would be tolerated. For example, a reasonable detection level of Dark matter depends on the energy scale associated with its mass and the background thermal noise. The presence of impurities also increases the importance of an anisotropic sensor where yearly and daily modulation of the DM detection rate provide a way to distinguish the DM signal from the background noise \cite{doi:10.1142/S0218271817300129}.

The focus of this paper is the effects of mass term fluctuations in DM on the electronic spectrum and the constraints one needs to meet to detect Dark matter using massed DM. As such the work presented here builds on the previous results. The effect of disorder on the vanishing density of states of the massless 2D and 3D Dirac point has been studied in detail \cite{PhysRevB.89.245110,Roy2016,PhysRevX.6.021042,doi:10.1146/annurev-conmatphys-033117-054037,PhysRevB.97.024204,Rostami2017,PhysRevB.93.085426}.

This paper is organized as follows.  Section~\ref{sec:gapped_dm_for_dm} elaborates on the relevant energy scales of the sensors fo Dark matter and how DM can be used as sensors. Section~\ref{sec:model} introduces the model of a Dirac cone with normal-distributed $m$-term. This model is used in Section~\ref{sec:interband_absorption} to calculate transition rates depending on material parameters. Section~\ref{sec:lifshitz_tails} extends the notion of Lifshitz tails to Dirac impurities. In Section~\ref{sec:competing_transitions} we show how thermal noise can interfere with the detection of dark matter. Finally, we summarize the results in Section~\ref{sec:conclusion}. We use natural units $\hbar=c=k_B=1$ unless specified otherwise.

\section{Gapped Dirac Materials and Dark Matter detection}
\label{sec:gapped_dm_for_dm}
To motivate the investigation of the disorder effects on small gap Dirac materials we start with the case of small mass Dirac material as a sensor material for light dark matter detection. The original idea to use Dirac materials for Dark matter (DM) detection was made by~\cite{PhysRevD.97.015004} who also identified a number of 3D inorganic compounds as candidate materials. The utility of small mass Dirac material follows from the following estimates. Assuming dark matter particles with MeV scale mass and given the velocity $v_\text{DM}\sim10^{-3}c$ of the halo of DM in the Galaxy~\cite{PhysRevD.97.015004}, the kinetic energy $\frac{1}{2}m_\text{DM}v^2_\text{DM}$ is on order of eV. Lighter dark matter particles (sub-MeV) require an energy scale of meV-eV. In this regime, gapped Dirac materials with a gap of $\mathcal{O}$(meV) provide the right energy scale for sensors. On the other hand, a small but finite gap is necessary to suppress thermal fluctuations in the  electric circuit and enable detection of dark matter. It is this tension between small mass gap desired for detection and yet finite gap required to reduce noise, that defines the search space for the Dirac material detectors for DM.

The density of dark matter in our galaxy is approximately 0.4 GeV/cm$^3$~\cite{doi:10.1142/S0218271817300129}. In the case of light dark matter $m_\text{DM}=\SI{100}{\kilo\electronvolt}$ (kinetic energy of $\SI{50}{\milli\electronvolt}$), this corresponds to a particle density of \SI{4000}{\per\cubic\centi\meter}. For $m_\text{DM}=\SI{1}{\mega\electronvolt}$ dark matter, the particle density is \SI{400}{\per\cubic\centi\meter}.

The 3D inorganic compounds ZrTe5 and ZrSe5 have been proposed as candidates for dark matter sensors. ZrTe5 has a experimental band gap of \SI{23.5}{\milli\electronvolt} and an anisotropic cone with Fermi velocities ranging $10^{-4}\sim 10^{-3}$~\cite{PhysRevD.97.015004}. A recently proposed organic candidate sensor material is (BEDT-TTF)$\cdot$Br with a theoretically estimated band gap of \SI{50}{\milli\electronvolt} and a Fermi velocity of $5 \times 10^{-4}\ c$~\cite{doi:10.1002/pssr.201800293}. Alternatively, 2D sensors have been proposed, such as graphene sheets~\cite{Hochberg2017}.

In all of these studies the effect of impurities and fluctuations on dark matter detection was not discussed in detail. Here we expand the approach and focus on disorder effects. 

\section{The model}
\label{sec:model}

\begin{figure}[t]
    \centering
    \includegraphics[width=\linewidth]{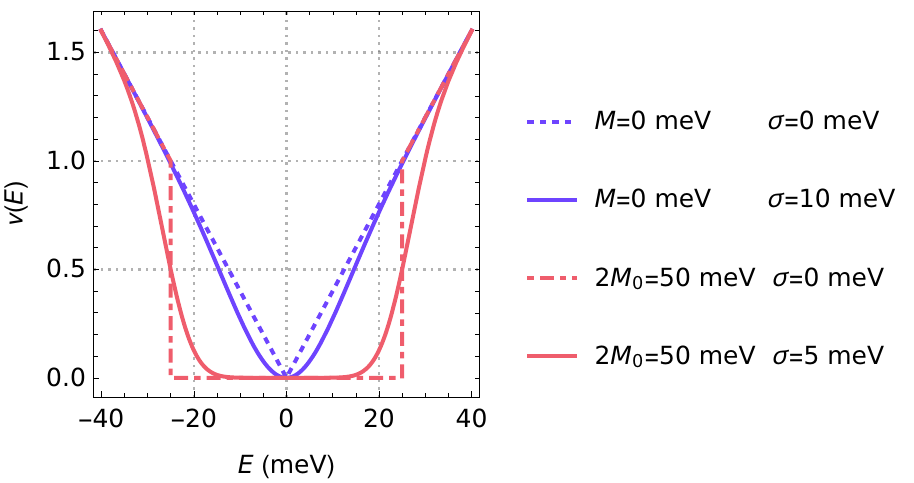}
    \caption{Density of states for the gapless ($M_0=0$) and gapped ($2M_0=\SI{50}{\milli\electronvolt}$) with standard deviation $\sigma$ in $M$. The tail of DOS near $E = 0$ needs to be taken into account in a discussion for realistic DM for sensor applications.}
    \label{fig:dos_curves}
\end{figure}

As explained, we model the effects of disorder with a random mass term. Consider a massive Dirac cone with a gap of $\Delta=2M$, described by the following stationary Hamiltonian and eigenvalues,
\begin{align}
    H&=v_F\ \bm{k}\cdot \bm{\sigma} + M\sigma_z\label{eq:hamiltonian}\\
    E_k&=\pm \sqrt{v_F^2 k^2 +M^2}.
\end{align}
The density of states of this 2D system is shown in Figure~\ref{fig:dos_curves} and has the form
\begin{align}
\nu(E)&=\int_{\mathbb{R}^2}\frac{d^2k}{(2\pi)^2}\sum_{n}\delta\left(E-E_k\right)\nonumber\\
    &=\frac{E}{2\pi v_F^2}\left(\theta\left(E+|M|\right)-\theta\left(E-|M|\right)\right).
\end{align}
We assume no long range correlations in physical mechanism opening the gap, and assume that randomness in mass term $M$ is normal-distributed due to the central limit theorem.

For sensor applications the Fermi velocity $v_F$ and mass-term $M$ are two key tuning parameters. The gap is tuned with $M$ to eliminate thermal noise but to still provide available states for small energy excitations. The Fermi velocity $v_F$ determines the slope of the Dirac cone and a shallow slope provides more available states.

The mass is randomly distributed around $M_0$ with a normal distribution $P(M)=\frac{1}{\sqrt{2\pi \sigma^2}}e^{-\frac{(M-M_0)^2}{2\sigma^2}}$ with the standard deviation $\sigma$. We obtain the averaged density of states,
\begin{align*}
    \left<\nu(E)\right>&=\int_{-\infty}^\infty dM\ P(M)\nu(E,M)\\
    &=\frac{E}{4\pi v_F^2}\left(\erf{\left(\frac{M_0+E}{\sqrt{2}\sigma}\right)}-\erf{\left(\frac{M_0-E}{\sqrt{2}\sigma}\right)}\right).
\end{align*}
Figure~\ref{fig:dos_curves} shows the density of states for a number of different scenarios (with and without disorder). Here we allow the mass term to take both positive and negative values due to the mass fluctuations. Technically, a model of disorder with the integration limited to positive $M$ is also possible and would lead to similar estimates.  The averaged density of states $\left<\nu(E)\right>$ vanishes at $E=0$ quadratically but remains finite for any non-zero $E$. These tails are compared to the case of a discrete random variable $M$, i.e. Lifshitz tails, in the next section.

\section{Lifshitz tails}
\label{sec:lifshitz_tails}
Lifshitz tails were proposed to account for the optical absorption of light with frequencies that were significantly lower than the original band gap in semiconductors \cite{lifshitz1988introduction}. The central point is the role of rare fluctuations where the probability of event can be exponentially small yet the significance of these small fluctuations is large. One example would be a tail in the density of states due to large regions in a disordered system where dopant atoms form a local crystal structure of a different kind than the surrounding crystal. Statistically, the probability for these regions is exponentially small and controlled by a large entropy fluctuations required to produce these regions $ P \sim \exp^{-S} $. Yet these regions often control the optical and transport properties at much softer energy scale than the surrounding crystal. For example, these rare regions can have massless excitations compared to the fully gapped bulk states. Hence one would tend to ignore the exponentially small probability of these events. Yet these exponentially unlikely regions control the low energy DOS in a fully gapped semiconductors.

We now extend the estimates typically used in semiconductors to gapped DM and estimate the role of Lifshitz tails. The rare regions in a $d$-dimensional semiconductor lead to $\left<\nu\right> \approx \exp{\left(\lambda E^{-d/2}\ln{c}\right)}$~\cite{0038-5670-7-4-R03}. For a finite impurity concentration $c$ there is a small but finite probability the entire sample is made up of impurity atoms and hence hosts a finite number at very low $E$. In the case of impurities inducing massless Dirac states $E=kv_F$ with concentration $c$, the density of states, 
\begin{align}
\left<\nu(E,c)\right>&\sim c^{V(E)}\nonumber\\
&= c^{\lambda E^{-d}},\quad\lambda = \left(\kappa_0 v_F\right)^{d}\nonumber\\
&=\exp{\left(\lambda E^{-d}\ln{c}\right)}
\end{align}
where coefficient $\kappa_0$ corresponds to the shape of volume $V=L^d$ with the lowest energy. Beyond the fact that the density of states of both the normal distributed $m$ model and the Lifshitz approach show tails, the tails are different. The DOS for Lifshitz tails decays exponentially towards $E\rightarrow 0$ whereas the tail in the random-mass model decays quadratically, as we show below.

In the semiconductor industry, expensive electronic-grade Silicon has impurity concentrations as low as $c\sim 10^{-10}$\cite{padovani1978process}. Cheaper metallurgical-grade Silicon has an impurity concentration of $c\sim 0.02$. The impurity concentrations that are present in synthesized candidate materials (such as ZrTe5, ZrSe5, (BEDT-TTF)$\cdot$Br) are unknown. We assume they could be on the range of few percent, consistent with $c\sim 0.02$ in regular silicon. 

The Lifshitz tails presented above treat the mass term as a discrete random variable where $P(0)=c$ and $P(M_0)=(1-c)$. The mass statistics $\left<M\right>=(1-c)M_0$ and $\left<M^2\right>=(1-c)M_0^2$ lead to the mass variance $\sigma^2=\left<M^2\right>-\left<M\right>^2=(c-c^2)M_0^2$. This estimate also  provides a link between the continuous random mass model and the discrete Lifshitz tails.

\section{Interband absorption}
\label{sec:interband_absorption}
The transition rate of the sensor material depends on  the material parameters: Fermi velocity $v_F$, mass $M$, mass standard deviation $\sigma$ and temperature $T$. Fermi's golden rule gives the transition rate from a state $\ket{i}$ to $\ket{f}$ due to the absorption of a photon with energy $\hbar\omega$. We consider minimal coupling $\bm{p}\rightarrow \bm{p}-q\bm{A}(t)$ in Equation~\ref{eq:hamiltonian}, leading to a time-dependent perturbation,
\begin{align}
V&=-v_F q\ \bm{\sigma}\cdot\bm{A}(t)\nonumber\\
&\approx - v_F q\ \bm{\sigma}\cdot\bm{A}_0\frac{1}{2}\Big(e^{i\omega t}+e^{-i\omega t}\Big),
\end{align}
where the vector potential $\bm{A}(t)$ is approximated by the dipole approximation and is assumed a linearly polarized plane wave of frequency $\omega$ \cite{PhysRevB.97.195123}. The matrix element of the perturbation between the initial and final states can be expressed as
\begin{align}
    \mathcal{M}&=-\frac{1}{2}v_Fq\braket{f|\bm{\sigma}\cdot \bm{A}_0|i}\nonumber\\
    \left|\mathcal{M}\right|^2&=\frac{q^2A_0^2v_F^2}{8E_k^2}\Big(2M^2+k^2v_F^2-k^2v_F^2\cos{\left(2(\alpha-\beta)\right)}\Big),
\end{align}
with polar angles $\alpha$ and $\beta$ for $\bm{A}_0$ and $\bm{k}$, respectively. Using the Fermi golden rule with the periodic perturbation $V$ we obtain the probability for a transition per unit time \cite[\S42]{LLV3},
\begin{align}
    \mathcal{R}(\omega)&=2\pi\int_{\mathbb{R}^2}d \bm{k}\ |\mathcal{M}|^2\delta\left(E_f(k)-E_i(k)-\omega\right)\nonumber\\
    &=2\pi \int_{0}^{\infty} d k\ k\ \delta\left(2E_k-\omega\right) \int_{0}^{2\pi}d\beta\ |\mathcal{M}|^2\nonumber\\
    &=\frac{A_0^2 \pi^2 q^2}{8\omega}\left(4M^2+\omega^2\right)\theta\left(\omega-2|M|\right)
    \label{eq:R}
\end{align}
As expected, transitions are forbidden within the $2M$ gap and any thermal noise with this energy scale is blocked. However, this changes when mass fluctuations introduce states in the gap as shown in the next section. Note that the transition rate does not depend on $v_F$. At the gap boundary of $\omega=2M$ the transition rate derivative is zero. For large $\omega$ the transition rate $\mathcal{R}$ scales linearly with slope $\frac{1}{8} A_0^2\pi^2 q^2$.

\subsection{Normal-distributed M}
\begin{figure}[t]
    \centering
    \includegraphics{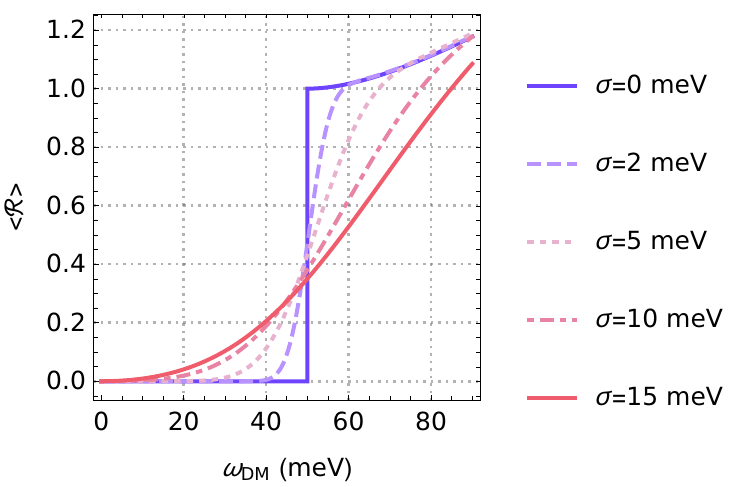}
    \caption{The effect of randomness in $M$ on the probability of a transition per unit time with a mass term $M_0=\SI{25}{\milli\electronvolt}$. Normalized such that $\left<\mathcal{R}(2M_0)\right>=1$ for $\sigma=\SI{0}{\milli\electronvolt}$.}
    \label{fig:R_curves}
\end{figure}
To include the effect of mass fluctuations on the transition rate $\mathcal{R}(\omega)$, we again consider the mass-term to be normally distributed around $M_0$ with standard deviation $\sigma$. This leads to the averaged transition rate,
\begin{widetext}
\begin{align}
    \left<\mathcal{R}(\omega)\right>&=\int_{-\infty}^{\infty}dM\ P(M)\mathcal{R}(\omega)\nonumber\\
    &=\frac{1}{16\omega}A_0^2\pi^{\frac{3}{2}}q^2 \Bigg(
    \sqrt{\pi} \left(4 \left(M_0^2+\sigma ^2\right)+\omega ^2\right)
   \left[
    \erf{\left(\frac{\Delta_+}{\tilde{\sigma}}\right)}-\erf{\left(\frac{\Delta_-}{\tilde{\sigma}}\right)}\right]
    -
    \tilde{\sigma} 
   e^{-\frac{\Delta_+^2}{8 \sigma ^2}} \left(\Delta_+
   e^{\frac{M_0 \omega }{\sigma ^2}}-\Delta_{-} \right)\Bigg),
    \label{eq:arif}
\end{align}
\end{widetext}
with $\Delta=2M_0$, $\Delta_+=\Delta+\omega$, $\Delta_{-}=\Delta-\omega$ and $\tilde{\sigma}=2\sqrt{2}\sigma$. Without randomness (i.e. $\sigma=0$) transitions are forbidden within the $\Delta$ gap. Mass fluctuations lead to a finite number of available states inside the gap for any non-zero $E$, with the Density of States at low energies near Dirac point scaling quadratically, leading to a finite transition rate, see Figure~\ref{fig:R_curves}.

\subsection{Temperature dependence with Fermi-Dirac statistics}
\begin{figure}[b]
    \centering
    \includegraphics[width=\linewidth]{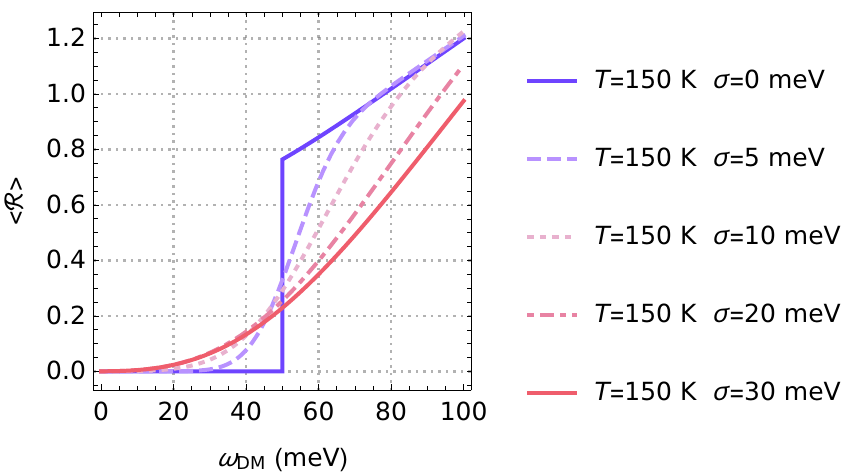}
    \includegraphics[width=\linewidth]{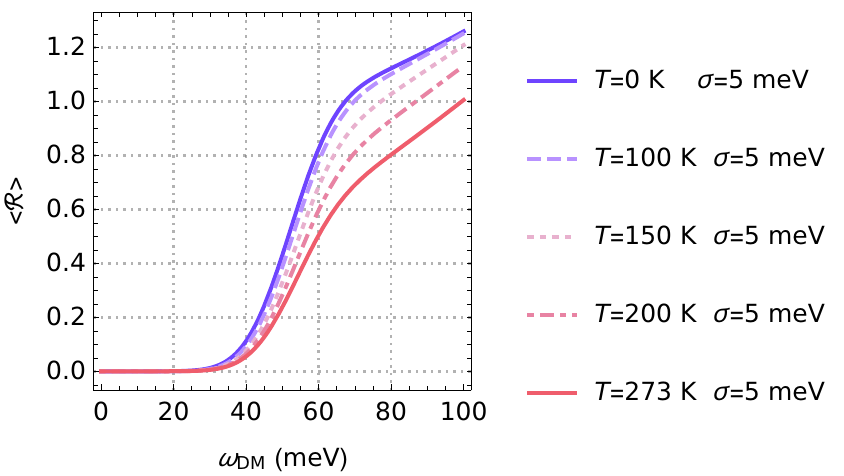}
    \caption{The probability of a transition per unit time $\mathcal{R}$ is reduced by an increase in temperature $T$ and smeared by the random mass $M$ distribution with standard deviation $\sigma$.}
    \label{fig:TA_curves}
\end{figure}
The transition rate discussed so far has no dependence on temperature. We can obtain the density of occupied states $n(E)=\left<\nu(E)\right>f(T,E)$ and density of holes with $p(E)=\left<\nu(E)\right>(1-f(T,E))$ where $f$ is the Fermi-Dirac distribution. Introducing temperature dependence to the Fermi golden rule is done by multiplying with the Fermi-Dirac distribution,

\begin{align}
    \mathcal{R}(\omega,T)&=2\pi\int_{\mathbb{R}^2}d \bm{k}\ |\mathcal{M}|^2\delta\left(2 E_k-\omega\right)f(E_i)\Big(1-f(E_f)\Big)\nonumber\\
    &=f\left(-\frac{\omega}{2}\right)\Big(1-f\left(\frac{\omega}{2}\right)\Big)\mathcal{R}(\omega)\\
    \left<\mathcal{R}(\omega,T)\right>&=f\left(-\frac{\omega}{2}\right)\Big(1-f\left(\frac{\omega}{2}\right)\Big)\left<\mathcal{R}(\omega)\right>
\end{align}
The results for a range of temperatures $T$ and mass standard deviation $\sigma$ are shown in Figure~\ref{fig:TA_curves}. Increasing temperature $T$ lowers the transition rate as unoccupied states in the conduction band become occupied.

\section{Competing transition rates and temporal modulations}
\label{sec:competing_transitions}
\begin{figure}[b]
    \centering
    \includegraphics{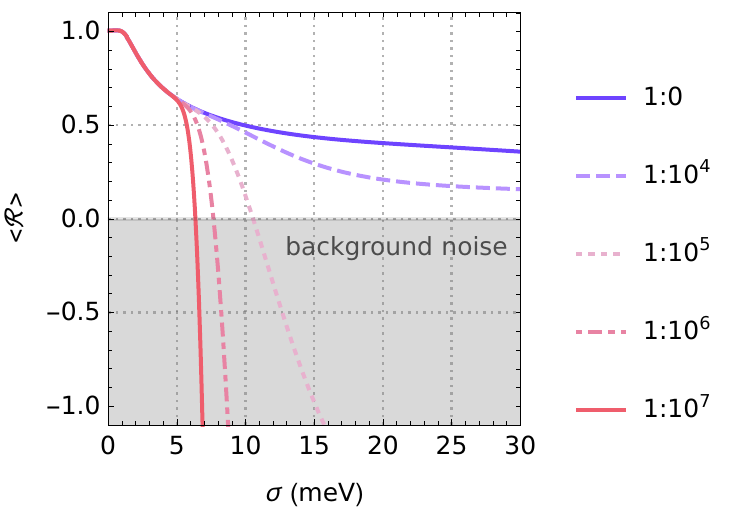}
    \caption{Competing transition rates $\left<\mathcal{R}\right>=A\left<\mathcal{R}\left(\omega_\text{DM}\right)\right>-B\left<\mathcal{R}\left(\omega_\text{th}\right)\right>$ for different ratios $A:B$ and increasing mass fluctuations $\sigma$. The transition rates assuming a mass-term of $M_0=\SI{25}{\milli\electronvolt}$ and $\omega_\text{DM}=\SI{55}{\milli\electronvolt}$ are compared to the transition rates due to thermal noise $\omega_\text{th}=\SI{0.4}{\milli\electronvolt}$. The shaded area indicates the range with more transitions due to noise than the photons of interest, i.e. $\left<\mathcal{R}\left(\omega_\text{th}\right)\right>>\left<\mathcal{R}\left(\omega_\text{DM}\right)\right>$.}
    \label{fig:signal_std}
\end{figure}
We can calculate and compare the transition rate due to the photons of interest $\left<\mathcal{R}\left(\omega_\text{DM}\right)\right>$ (i.e. dark matter) and the rate due to the background noise $\left<\mathcal{R}\left(\omega_\text{th}\right)\right>$. Background thermal noise at \SI{4}{\kelvin} (liquid Helium) in the range of $\SI{0.4}{\milli\electronvolt}$ are not measured without impurities (i.e. $\sigma=0$), because it is well within the $2M_0=\SI{50}{\milli\electronvolt}$ band gap. However, the presence of any mass fluctuations, i.e. $\sigma>0$, will introduce transitions due to the background noise.

Figure~\ref{fig:signal_std} shows a comparison of the transition rates by subtracting the rate induced by thermal noise from the transition rate by the dark photons of interest,
\begin{equation}
\label{eq:rate_comparison}
    \left<\mathcal{R}\right>=A\left<\mathcal{R}\left(\omega_\text{DM}\right)\right>-B\left<\mathcal{R}\left(\omega_\text{th}\right)\right>,
\end{equation}
where $A:B$ represents the ratio of DM and thermal (background) photons.
For example, for a $1:10^5$ ratio of DM and background we notice that at $\sigma=\SI{10}{\milli\electronvolt}$ the thermal transition rate takes over and DM photons are beyond the limit of detection.

For a given impurity concentration and thermal noise, we can estimate the upper bound on the DM/background noise ratio in order to remain sensitive to dark matter. We give two estimates based on above analysis: 

\begin{enumerate}[label=\alph*)]
\item A Dirac sensor with impurity concentration of $c\sim 0.05$ and a mass of $M_0=\SI{25}{\milli\electronvolt}$ has random mass fluctuations of $\sigma=\sqrt{(1-c)cM_0^2}\sim\SI{5.4}{\milli\electronvolt}$. Assuming background thermal noise of $\omega_\text{th}=\SI{0.4}{\milli\electronvolt}$ (liquid Helium at \SI{4}{\kelvin}), we can compare the transition rate due to dark matter $\left<\mathcal{R}(\omega_\text{DM})\right>$ with the background noise $\mathcal{R}(\omega_\text{th})$. For a DM/thermal ratio lower than $1:10^{8}$, the total rate $\left<\mathcal{R}\right>>0$ (Equation~\ref{eq:rate_comparison}), i.e., the DM rate $\left<\mathcal{R}(\omega_\text{DM})\right>$ is larger than the background noise which is necessary to be sensitive to dark matter.  

\item Considering an impurity concentration of $c\sim 0.02$ (a significantly cleaner sample) and gap  $M_0=\SI{25}{\milli\electronvolt}$, the random mass fluctuation is $\sigma\sim \SI{3.5}{\milli\electronvolt}$. Again, assuming thermal noise of $\omega_\text{th}=\SI{0.4}{\milli\electronvolt}$, a DM/background ratio lower than $1:10^{14}$ is necessary to be sensitive to dark matter. Given the estimate of dark matter density \SI{4000}{\per\cubic\centi\meter} (see Section~\ref{sec:gapped_dm_for_dm}) and the DM/background ratio above, this corresponds to an upper limit on the background noise density of $4000\cdot 10^{14}\sim\SI{E17}{\per\cubic\centi\meter}$.
\end{enumerate}

We now comment on temporal modulations of the rates.  The dark matter "wind" moves in random directions whereas the Earth is moving around the sun, which should cause annual modulation in the dark matter detection rate~\cite{PhysRevD.33.3495}. Additionally, the detector rotates daily around the Earth's axis. If the sensor material possesses an anisotropic dispersion, the DM signal should show daily modulation~\cite{Hochberg2017,2019arXiv190909170C}.
The transition rate (Equation~\ref{eq:R}) in the case of an anisotropic Dirac cone becomes,
\begin{equation}
    \mathcal{R}(\omega)=\frac{\pi^2q^2 \left(|\bm{A}_0\odot \bm{v}|^2\right)}{8 v_x v_y \omega}\left(4M^2+\omega^2\right)\theta(\omega-2|M|)
\end{equation}
where $\odot$ denotes the Hadamard product. In the case of an isotropic $A_0$, the transition rate includes a $\frac{v^2}{v_xv_y}$ prefactor. In this case, for a sensor with  $v_x/v_y=\eta(t)$ (due to Earth's rotation), the modulation would introduce a prefactor $\left(\eta(t)+1/\eta(t)\right)$ to the transition rate. Thus the highly anisotropic Dirac cones would significantly increase the daily modulations, e.g, for an anisotropy parameter $\eta$ ranging on the order of $\eta \sim 10$ the modulation will increase proportionately. We thus propose that highly anisotropic DM would be a better candidate materials for detection of annual modulation. 

\section{Conclusion}
\label{sec:conclusion}
We discussed the   fluctuations of the gap-opening mass term $m$ and rare fluctuations in gapped Dirac materials, which we model as Lifshitz tails or a random mass fluctuations. These fluctuations are naturally expected in realistic materials and would lead to constraints and tolerances allowed in gapped Dirac materials used for Dark matter detection. Fluctuations lead to a softening of the gap and significant gap filling with impurity tails.

We demonstrate the effect of randomness of the mass term in $2$-dimensional gapped Dirac materials on the transition rates due to interband absorption. The material parameters Fermi velocity $v_F$, mass $M$, mass fluctuations $\sigma$ and temperature $T$ should be tuned according to the specific sensor application. 

The proposed analysis is also relevant in the broad context of disorder and thermal effects in optical transitions in disordered DM.  Our results further extend the notion  of Lifshitz tails and random mass fluctuations to Dirac materials.

\section{Acknowledgement}\label{acknowledgements}
We are grateful to R.~M.~Geilhufe, D.~N.~Carvalho, J.~H.~Bardarson, J.~Conrad  and A.~Ferella for useful discussions. This work is supported  by the Swedish Research Council Grant,  the VILLUM FONDEN via the Centre of Excellence for Dirac Materials (Grant No. 11744) and the European Research Council ERC HERO grant.

\bibliography{references}

\end{document}